\documentclass[twocolumn,showpacs,preprintnumbers,amsmath,amssymb,aps,prb]{revtex4}

\usepackage{graphicx}
\usepackage{dcolumn}
\usepackage{bm}
\usepackage{subfigure}
\usepackage{natbib}
\usepackage{multirow}

\begin{document}

\title{Time resolved photoemission spectroscopy of electronic cooling and localization in CH$_3$NH$_3$PbI$_3$ crystals}

\author{Zhesheng Chen$^{1}$, Min-i Lee$^{2}$,  Zailan Zhang$^{3}$, Hiba Diab$^{4}$, Damien Garrot$^{5}$, Ferdinand L\'ed\'ee$^{4,6}$, Pierre Fertey$^{7}$, Evangelos Papalazarou$^{2}$, Marino Marsi$^{2}$, Carlito Ponseca$^{8}$, Emmanuelle Deleporte$^{4,6}$, Antonio Tejeda$^{2}$ and Luca Perfetti$^{1}$}

\affiliation{$^{1}$ Laboratoire des Solides Irradi\'{e}s, Ecole polytechnique,e, CNRS, CEA, Universit\'e Paris-Saclay, 91128 Palaiseau cedex, France}
\affiliation{$^{2}$ Laboratoire de Physique des Solides, CNRS, Universit\'{e} Paris-Saclay, Universit\'{e} Paris-Sud, 91405 Orsay, France}
\affiliation{$^{3}$ Institut de Min\'{e}ralogie et de Physique des mat\'eriaux et de cosmochimie (IMPMC), UMR CNRS 7590, Universit\'{e} Pierre et Marie Curie - case 115, 4, place Jussieu, 75252 Paris cedex 05, France}
\affiliation{$^{4}$ Laboratoire Aim\'{e} Cotton, Ecole Normale Sup\'{e}rieure ENS Paris-Saclay, CNRS, Universit\'{e} Paris-Sud, Universit\'{e} Paris-Saclay, 91405 Orsay, France}
\affiliation{$^{5}$ Groupe d'Etudes de la Mati\`{e}re Condens\'{e}e (GEMaC), CNRS, Universit\'{e} de Versailles, Saint-Quentin-en-Yvelines, Universit\'{e} Paris-Saclay, 45 Avenue des Etats-Unis, 78035 Versaille Cedex, France}
\affiliation{$^{6}$ Laboratoire de Photophysique et Photochimie Supramol\'{e}culaires et Macromol\'{e}culaires de l'Ecole Normale Sup\'{e}rieure de Cachan, 61 Avenue du Pr\'{e}sident Wilson, 94235 Cachan, France}
\affiliation{$^{7}$ Soci\'{e}t\'{e} civile Synchrotron SOLEIL, L'Orme des Merisiers, Saint-Aubin - BP 48, 91192 GIF-sur-YVETTE, France}
\affiliation{$^{8}$ Division of Chemical Physics, Lund University, Box 124, 221 00 Lund, Sweden}

\begin{abstract}
We measure the surface of CH$_3$NH$_3$PbI$_3$ single crystals by making use of two photon photoemission spectroscopy. Our method monitors the electronic distribution of photoexcited electrons, explicitly discriminating the initial thermalization from slower dynamical processes. The reported results disclose the fast dissipation channels of hot carriers (0.25 ps), set a upper bound to the surface induced recombination velocity ($<4000$ cm/s) and reveal the dramatic effect of shallow traps on the electrons dynamics. The picosecond localization of excited electrons in degraded CH$_3$NH$_3$PbI$_3$ samples is consistent with the progressive reduction of photoconversion efficiency in operating devices. Minimizing the density of shallow traps and solving the aging problem may boost the macroscopic efficiency of solar cells to the theoretical limit.
\end{abstract}


\maketitle

\section{Introduction}

Hybrid metal-organic perovskite semiconductors such as methylammonium lead iodide CH$_3$NH$_3$PbI$_3$ (MAPbI3) have emerged as promising new materials for photovoltaic devices and optoelectronics. The power conversion efficiency of solar cells based on perovskite materials exceeded 20\% within eight years \cite{Zhou,Yang,Table} and a theoretical limit of 30\% has been proposed \cite{Wei}. Such an impressive performance results from favorable material properties such as direct band gap of roughly 1.6 eV, large absorption coefficient and high charge carriers mobility \cite{Brenner}.

The perovskite materials can be broadly divided into two types: bulk crystals and polycrystalline thin films. Single crystals of MAPbI3, have long electron-hole diffusion length (up to 175 $\mu$m) and low density of recombination centers \cite{Cao,Valverde,Tian}. Thin films are more suitable for device application but display photoexcited carriers with smaller mobility and shorter lifetime \cite{Ponseca,Herz_Reco,Herz_Moby}. Moreover, the fast degradation of thin films in operating conditions poses serious limits to viable applications. Both the photoexposure and annealing in atmospheric conditions induces the formation of PbI$_2$ inclusions\cite{Jemli}. The main objective of our work is to show that compositional disorder leads to shallow traps where carriers localize on the picosecond timescale.


The carriers diffusion, localization and recombination is investigated by Two Photon PhotoEmission (2PPE) experiments on single MAPbI3 crystals. Remarkably, 2PPE monitors the energy distribution of excited electrons with good temporal resolution and high surface sensitivity. Moreover, the clean surface of a single crystal is a model and well controlled system where to explore the impact of localized states on the electronic motion.

The article is organized as follow: Section II contains X-rays diffraction, photoluminescence and photoemission characterization of our crystals. Section III describes the methodology and the technical aspects of the 2PPE technique. Section IV discusses the subpicosecond cooling of excited electrons. Section V proposes a diffusion model to explain the observed evolution of 2PPE signal. Section VI investigates the carriers localization in intentionally degraded crystals. Section VII deals with carriers recombination and possible effects of the surface. Section VIII reports the conclusions and acknowledgments.

\begin{figure}
\includegraphics[width=1\columnwidth]{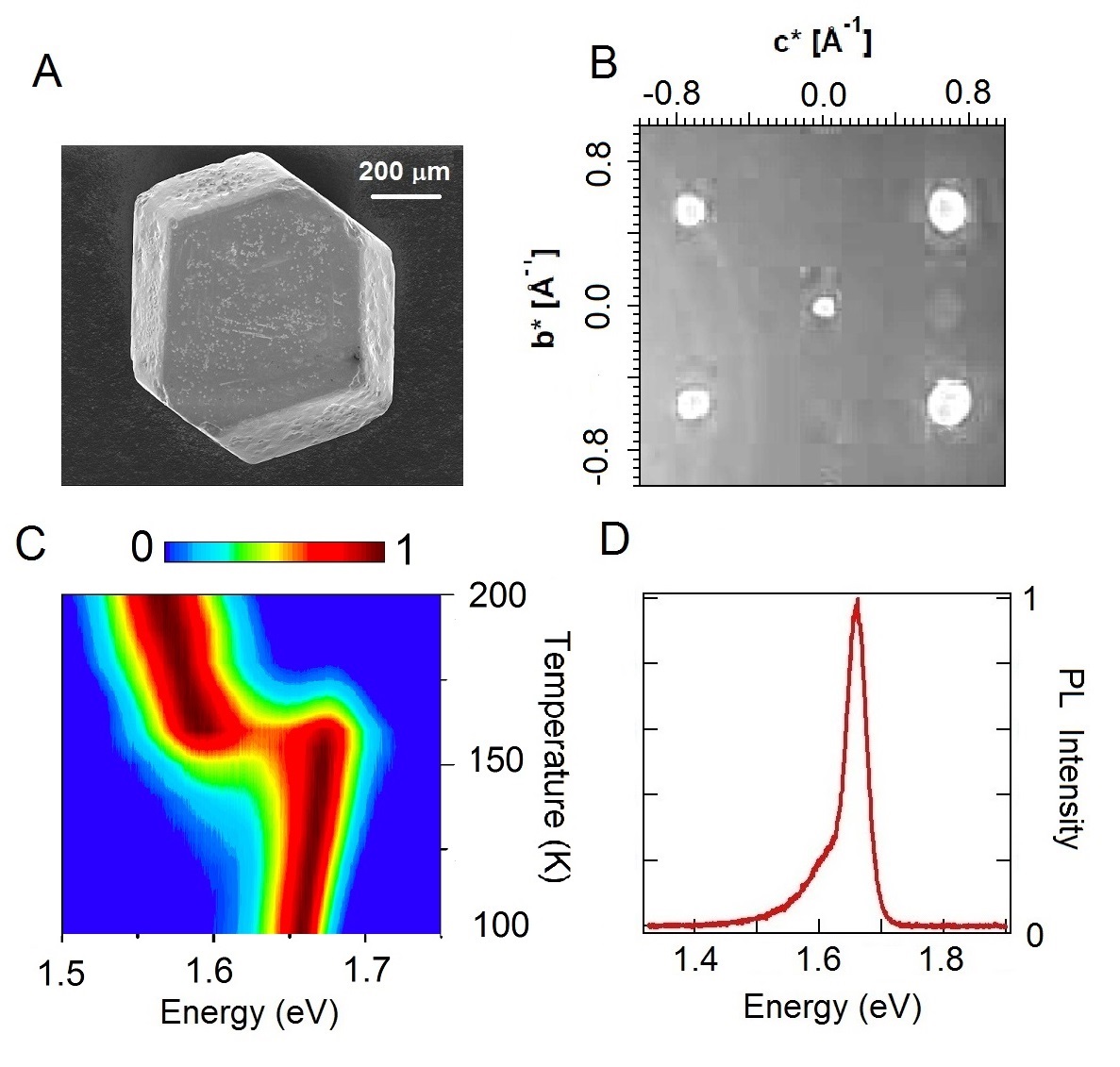}
\caption{A: Image of a single MAPbI3 crystal acquired by scanning electron microscope. B: X-ray diffraction of Bragg peaks in the \{0,$k_\bot$,$k_{||}$\} plane of the tetragonal phase. C: Intensity map of photolumiscence as function of photon energy and sample temperature. The abrupt transition around 160 K is due to the orthorhombic to tetragonal phase transition. D: Photoluminescence spectrum of MAPbI3 measured at 130 K.  }
\label{1M}
\end{figure}

\section{Sample characterization}

We investigate single crystals of MAPbI3 grown by inverse crystallization method.
Methylammonium iodide (0.78 g, 5 mmol) and lead iodide (2.30 g, 5 mmol) were dissolved in gamma-butyrolactone (5 mL) at 60 $^\circ$C. The yellow solution (2 mL) was placed in a vial and heated at 120 $^\circ$C during one to four hours depending on the desired crystal size. As shown in the scanning electron microscope image of Fig.  \ref{1M}A, the samples are crystals of millimetric size. 
X-ray diffraction measurents have been performed at the CRISTAL beamline of synchrotron SOLEIL by means of a four-circle diffractometer. The Bragg's reflections in Fig. \ref{1M}B confirm the high quality of the single crystals \cite{Antonio}.

Figure \ref{1M}C displays the photoluminescence map of the MAPbI3 single
crystals between 100 K and 200 K. The emission is composed of a single peak arising from electron-hole recombination across the band-gap. Upon cooling, the sudden blueshift of the emission line at $\cong160$ K \cite{Deleporte} is due to the widening of band gap at the tetragonal to orthorhombic phase transition. Below 160 K, the evolution of photoluminescence with temperature indicates a reduction of the band gap with lattice contraction. From the photoluminescence spectrum recorded at 130 K (see Fig. \ref{1M}D), we extract the band gap energy $\Delta_g=1.66$ eV. In the following, we will refer all spectroscopic measurements to the orthorhombic phase at 130 K.

\begin{figure}
\includegraphics[width=1\columnwidth]{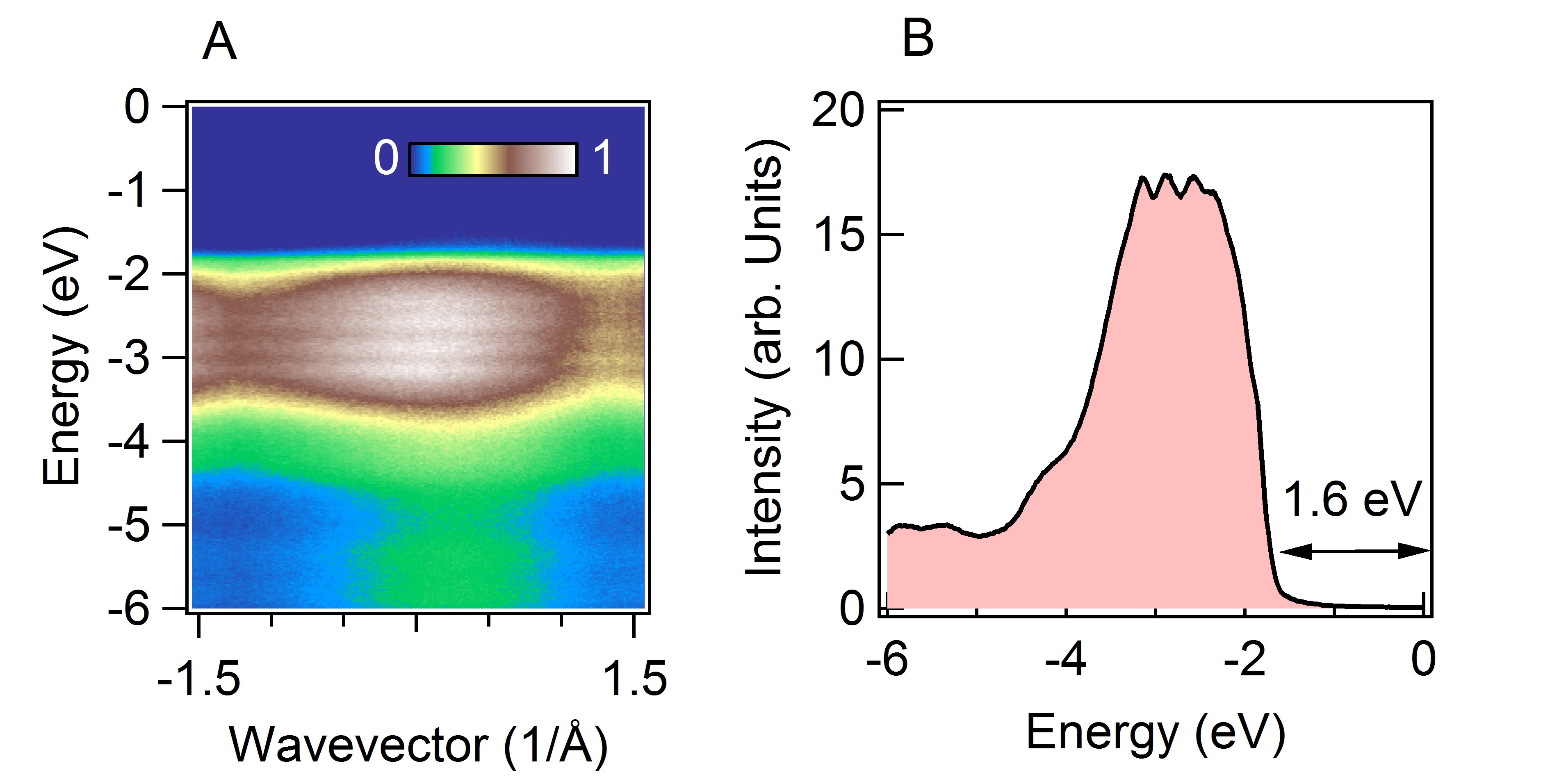}
\caption{A: Photoelectron intensity map acquired with photon energy of 94 eV in the  \{$k_{||}$,$k_\bot\sim 0$,0\} direction of the tetragonal phase. B: Wavevector integrated photoelectron spectrum acquired with photon energy of 94 eV. The maximum of the valence band is located 1.6 eV below the chemical potential.}
\label{2M}
\end{figure}

Single crytals have been mounted on the \{0,1,0\} plane and cleaved in ultra high vacuum at base pressure below 10$^{-10}$ mbar. Despite the high crystalline quality of our samples, Low Energy Electron Diffraction (LEED) did not display any Bragg spot. We evince that cleaved surfaces are rough, probably because of the brittle nature of MAPbI3. Figure \ref{2M}A shows Angle Resolved PhotoElectron Spectroscopy (ARPES) measurements performed at the CASSIOPEE beamline of synchrotron SOLEIL. The selected photon beam of 94 eV maximizes the cross section of the valence band and corresponds to a perpendicular wavevector $k_\bot\sim 0$. Variations of spectral intensity with respect to $k_{||}$ are consistent with the electronic band dispersion in the first Brillouin zone \cite{Antonio}. We show in Fig. \ref{2M}B the spectrum obtained by integrating  the photoelectron map in the interval [-1.5,1.5] \AA$^{-1}$. The chemical potential $\mu_F$ is located 1.6 eV higher than the top of the valence band, indicating that MAPbI3 crystals are naturally $n$-doped.

\begin{figure}
\includegraphics[width=0.7\columnwidth]{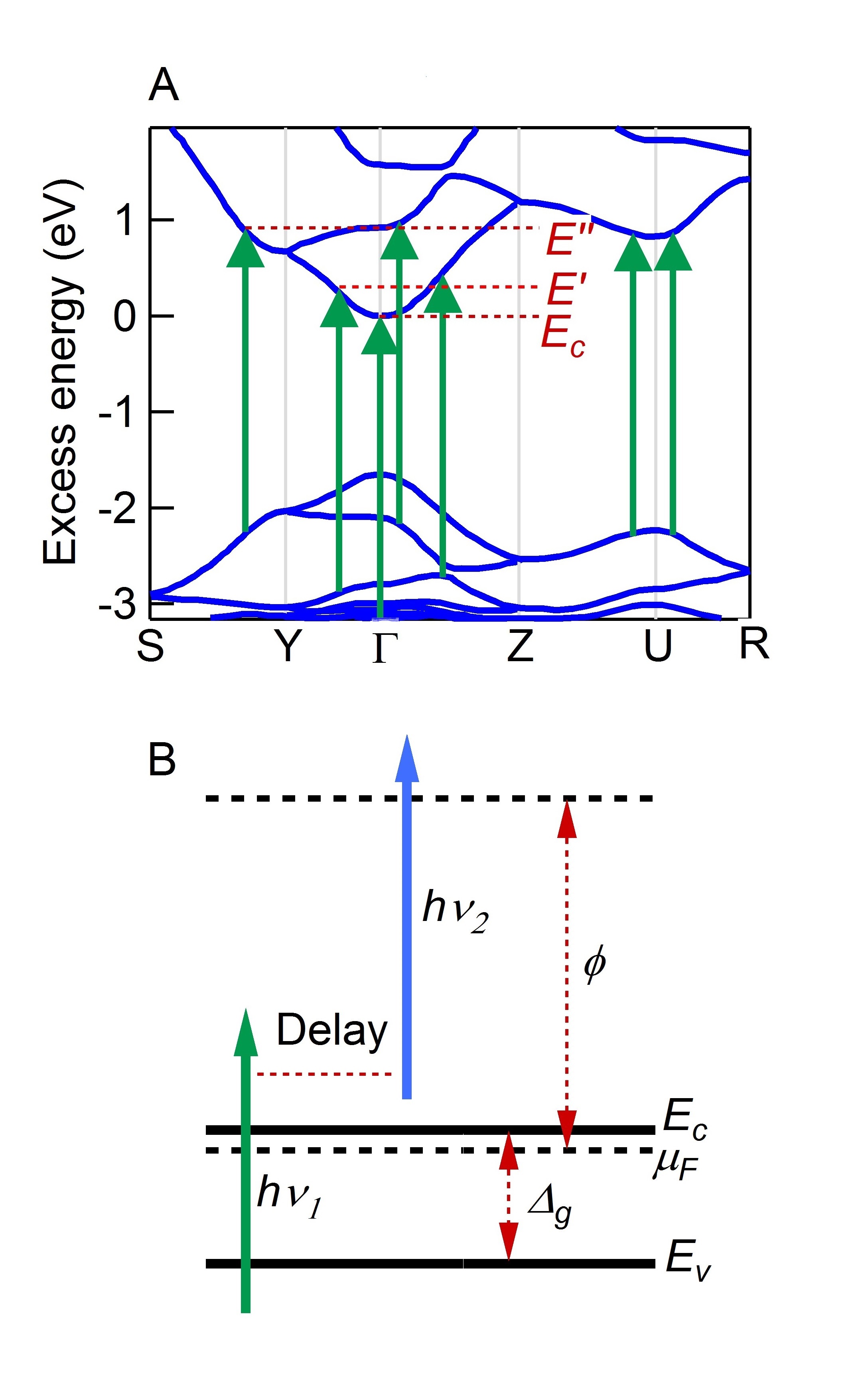}
\caption{A: Band structure of MAPbI3 freely adapted from Filip \textit{et al.} \cite{Filip}. The green arrows stand for direct transitions induced by photons with $h\nu_1= 3.15$ eV. B: Energetics of the 2PPE experiment with pump photon energy $h\nu_1$, probe photon energy $h\nu_2$, band gap $\Delta_g$, chemical potential $\mu_F$ and analyzer work function $\phi$.}
\label{3M}
\end{figure}

\section{Two Photon PhotoEmission}

The temporal evolution of the excited state is measured by means of Two Photon PhotoEmission (2PPE). Our photon source is a Ti:Sapphire laser system delivering 6 $\mu$J pulses with repetition rate of 250 kHz. Part of the fundamental beam ($h\nu_0 = 1.55$ eV) is converted to the second harmonic ($h\nu_1= 3.15$ eV) in a $\beta$-BBO crystal while the rest is employed to generate the third harmonic ($h\nu_2 = 4.7$ eV) \cite{Faure}. We photoexcite the sample at 130 K by 50 fs pump pulses of 20 $\mu$J/cm$^2$, centered at $h\nu_1$. According to the reported value of the absorption coefficient \cite{Green}, this pulse results in an electron-hole density of $8\times10^{18}$ cm$^{-3}$. 

As shown in Fig. \ref{3M}A the photons of the pump beam generate photoexcited electrons with excess energy up to $h\nu_1-\Delta_g=1.5$ eV, thereby inducing optical transitions in two branches of the conduction band. After a variable delay time, probe pulses centered at $h\nu_2=4.7$ eV promote the excited electrons above the vacuum level (see Fig. \ref{3M}B). Photoelectrons outgoing from the sample are detected by a hemispherical energy analyzer with an acceptance angle of roughly $5\times 1$ degrees$^2$ around normal emission. The overall energy resolution of 60 meV is dominated by the bandwidth of the probe beam. This technique provides a direct mapping of the electronic distribution in the photoexcited sample. Moreover, the electrons that have absorbed the two photons hold an ineastic mean free path of few nanometers \cite{Unal}. Such high surface sensitivity can be exploited to question the electrons dynamics in the topmost layers of the cleaved crystal.

\section{Ultrafast cooling of hot electrons}

\begin{figure}
\includegraphics[width=1\columnwidth]{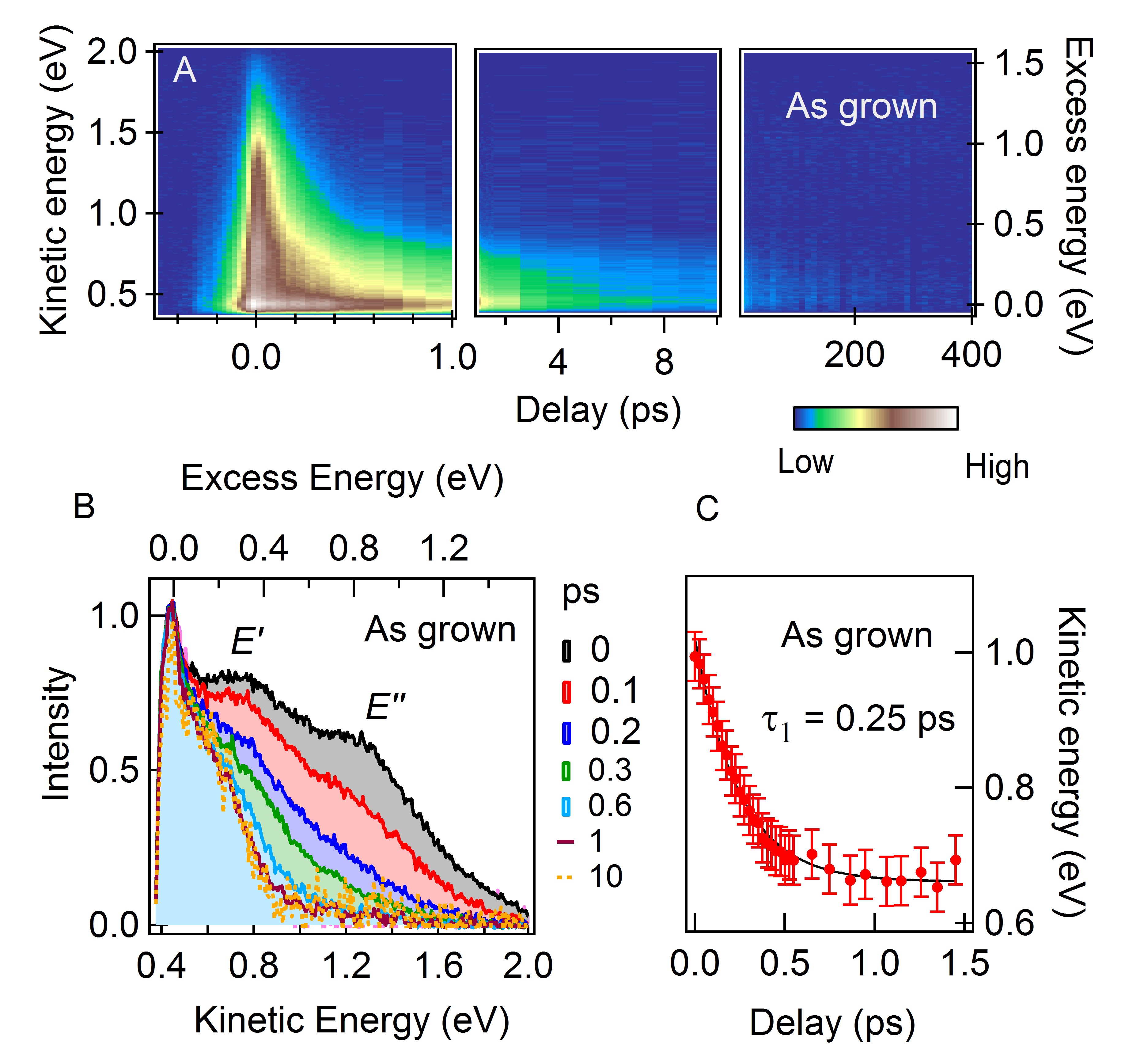}
\caption{A: Photoelectron intensity map in the as-grown sample as a function of kinetic energy and pump-probe delay. B: Energy distribution curves acquired at different values of the pump probe delay and normalized to their maximum value. C: Evolution of the average kinetic energy as a function of time. The solid line is an exponential fit with time constant $\tau_1=0.25$ ps.}
\label{4M}

\end{figure}

We show in Fig. \ref{4M}A a color scale plot of the photoelectron intensity acquired as a function of kinetic energy and pump probe delay. The nominal kinetic energy of the conduction band minimum is $E_c= h\nu_2-\phi+\Delta_g-\mu_F+E_v=0.45$ eV, where $\phi=4.3$ eV is the workfunction of our analyzer. As shown in Fig. \ref{4M}B, the spectrum acquired at the maximal overlap between pump and probe pulse (zero delay) displays a peak in proximity of $E_c$, a second peak near to $E'=E_c+0.35$ eV and a third peak at $E''=E_c+0.9$ eV. As sketched in Fig. \ref{3M}A, the structures $E'$ and $E''$ are located at excess energies where the 3.15 eV photons generate high density of optical transitions \cite{Filip}.

Next, we consider the dynamics of the electrons soon after the arrival of the pump pulse.
Figure \ref{4M}B shows that the excited electronic distribution varies strongly within the first picosecond. The electrons with large excess energy thermalize towards the bottom of the conduction band by means of electron-electron scattering and phonon emission. We exclude the occurrence of carriers multiplication because the photon energy of the pump beam is lower than twice the value of the bandgap. The initial thermalization time can be quantified by evaluating the average kinetic energy contained in the spectrum as a function of time. In practice, we calculate $\langle E\rangle_t= \int I(E,t) E dE/\int I(E,t) dE$, where $E$ is the kinetic energy, $I(E,t)$ is the photoelectron intensity, and $t$ is the pump-probe delay.

Figure \ref{4M}C shows the resulting $\langle E\rangle_t$ together with the fit by an exponential decay with time constant of $\tau_1=0.25$ ps. This value is in excellent agreement with recent 2PPE experiments on MAPbI3 thin films \cite{Niesner} and it is faster than the energy relaxation time observed in inorganic semiconductors \cite{Tanimura}. We suggest that the rapid electronic cooling of MAPbI3 arises via the coupling of excited carriers to the internal vibrations of the CH$_3$NH$_3^+$ cations. For example, the stretching modes of C-H and N-H bonds holds quantum energy of $370-400$ meV and can efficiently drain the excess energy of photoexcited electrons \cite{Brivio}. Moreover, the recent observation of optical sidebands in two-dimensional hybrid perovskites indicate a sizable coupling of the exciton to an organic mode with quantum energy of 40 meV \cite{Straus}.

According to Fig. \ref{4M}C, the hot electrons will reach quasi-equilibrium with the coupled phonons after $\cong 3\tau_1=0.75$ ps. Several authors suggested that strongly coupled modes attain an occupation level higher than the thermal one and relax on the slower timescale of anharmonic interaction \cite{Beard_hot,Price}. In this respect, time resolved Raman experiments may be helpful to address the specific vibrational modes where electrons transfer their excess energy \cite{Heinz}.
Note in Fig. \ref{4M}B that the spectra acquired at delay time of 1 ps and 10 ps display a shoulder at excess energy $\sim0.2$ eV. Niesner \emph{et al.} observed a similar shoulder only in the tetragonal phase of MAPbI3 and they ascribed it to an unconventional kind of polaronic dressing \cite{Niesner,Zhu}. This issue requires additional measurements and it is currently under investigation.

\section{Diffusion in the as-grown sample}

\begin{figure}
\includegraphics[width=\columnwidth]{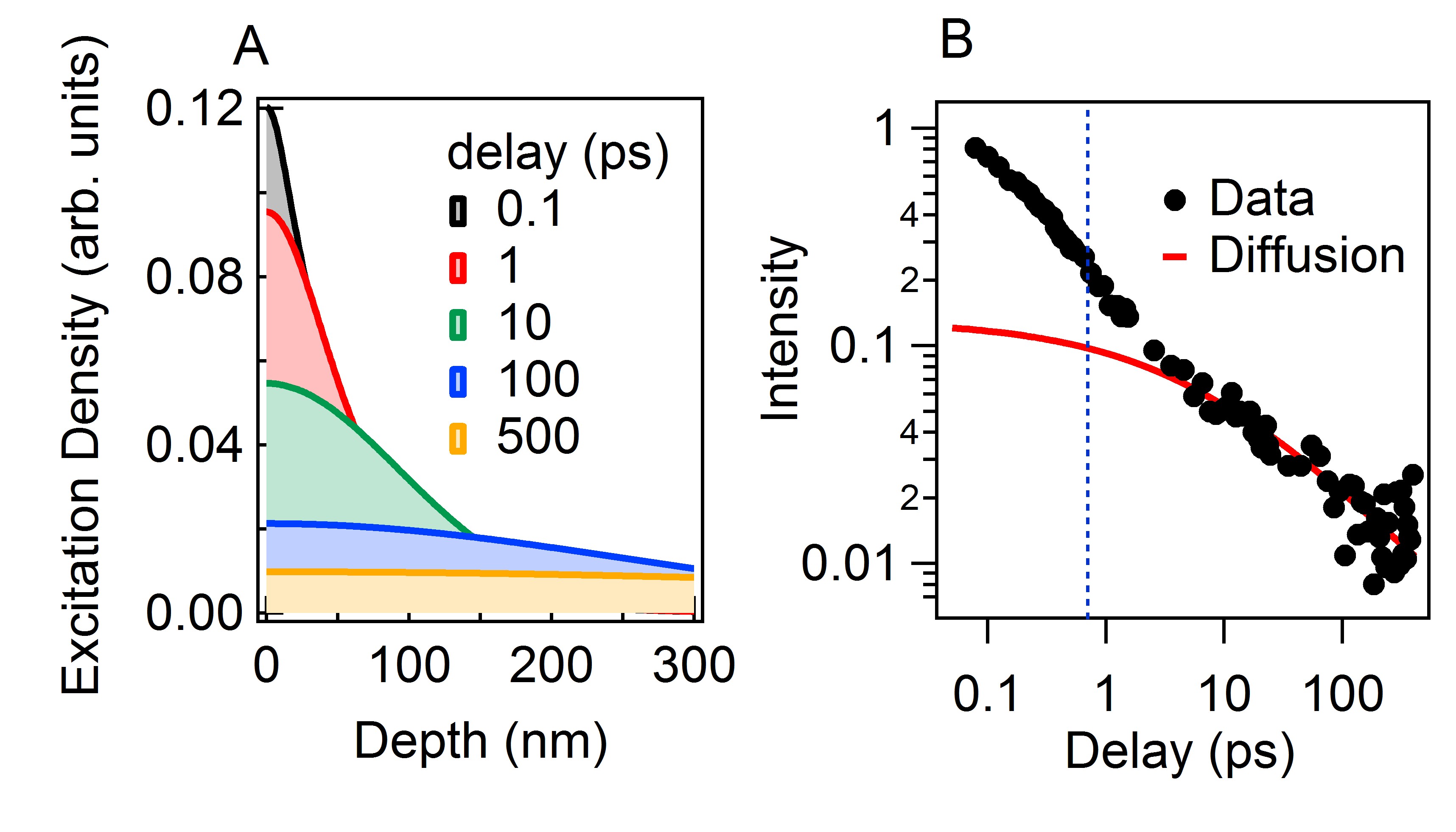}
\caption{A: The density profile of photoexcited electrons is calculated by the diffusion model of equation \ref{eq1} for selected delay times. B: The integrated intensity of the 2PPE signal (black marks) as a function of time is compared with the diffusion model (red line). The dotted blue line at $t=3\tau_1$ indicates the delay time when electrons have fully thermalized.}
\label{5M}
\end{figure}

Since the 2PPE technique probes carriers within the topmost nanometers, the integrated signal $\int I(E,t) dE$ follows the instantaneous electronic concentration at the surface of the sample times the photoemission cross section. Once the electrons have fully thermalized, the photoemission cross section becomes time independent. Therefore, the progressive decay of 2PPE signal that can be observed in the middle and right panel of Fig. \ref{4M}A is due to the drift-diffusion of electronic charges from the surface into the bulk. First, we analyze the effect of pure diffusion on the long timescale dynamics. Electrons that are excited in an optical penetration depth \cite{Green} $\alpha^{-1}=50$ nm move in to the bulk because of Brownian motion. The electronic concentration at distance $x$ from the surface and pump-probe delay $t$ is given by \cite{Beard_Reco}

\begin{eqnarray}
n(x,t)\propto \frac{1}{2}\exp\left(-\frac{x^2}{4Dt}\right)w\left(\alpha\sqrt{Dt}-\frac{x}{2\sqrt {Dt}}\right)+\nonumber\\
+\frac{1}{2}\exp\left(-\frac{x^2}{4Dt}\right)w\left(\alpha\sqrt{Dt}+\frac{x}{2\sqrt {Dt}}\right),
\label{eq1}
\end{eqnarray}

where $w(z)=\exp(z^2)(1-\mathrm{erf}(z))$ and $D$ is the electrons diffusion constant of MAPbI3 crystals at 130 K.   We recall that the mobility $\mu$ of MAPbI3 is limited by electron-phonon scattering \cite{Herz_Moby} and scales as $T^{-3/2}$. Therefore, we can invoke the Einstein relation $D=\mu k_bT\propto T^{-1/2}$ and refer to the literature value of $D$ at room temperature \cite{Cao} in order to estimate the electron diffusion constant at T=130 K. The resulting $D=3$ cm$^{2}$/s is plugged in to equation 1 of the electronic density. We show in Fig. \ref{5M}A the resulting $n(x,t)$ as a function of $x$ for selected values of $t$.
Figure \ref{5M}B compares the calculated $n(0,t)$ with the experimental evolution of the photoemission signal in the temporal window [0.05,400]  ps. The measured decay follows the diffusion behavior for delay time larger than 3 ps but proceeds much faster during the first picosecond. Niesner \emph{et al.} ascribed this initial drop to the rapid variation of the photoemission cross section during the thermalization process \cite{Niesner}. On one hand, it is fully plausible that carriers changing energy and wavevector display sudden changes of cross section. On the other hand, the measured intensity deviates from the diffusion model even after that the electrons have fully thermalized (namely for $3\tau_1<t<10\tau_1$). Eventually, built-in fields at the sample surface lead to an initial drift of the electrons out of the detection regions. Such ultrafast segregation between electrons and holes has been recently observed in other doped semiconductors \cite{Hajlaoui}. In the case of MAPbI3 the built-in field may originate from the residual polarity of the crystal termination \cite{She}.

\begin{figure}
\includegraphics[width=1\columnwidth]{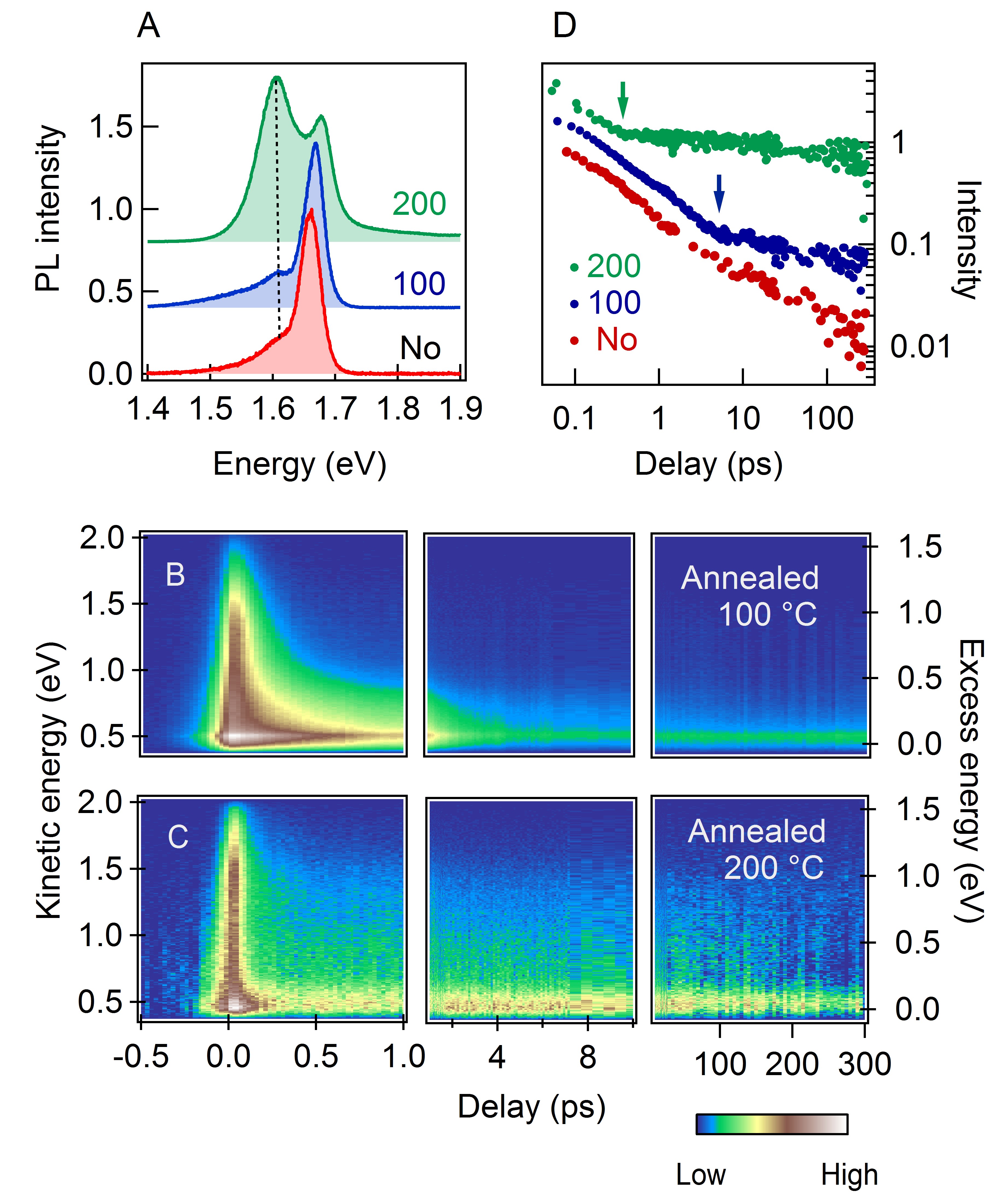}
\caption{A: Photoluminescence spectra acquired as grown, 100 $^\circ$C annealed and 200 $^\circ$C samples. The dashed line is a guide to the eye showing the development of trapped states upon annealing. Photoelectron intensity map in samples annealed at 100 $^\circ$C (panel B) and 200 $^\circ$C (Panel C). D: Temporal evolution of the integrated 2PPE signal in the pristine and annealed samples. The arrows indicate the characteristic timescale when electronic trapping takes place.}
\label{6M}
\end{figure}

\section{Electronic localization in annealed samples}

\begin{figure}
\includegraphics[width=0.75\columnwidth]{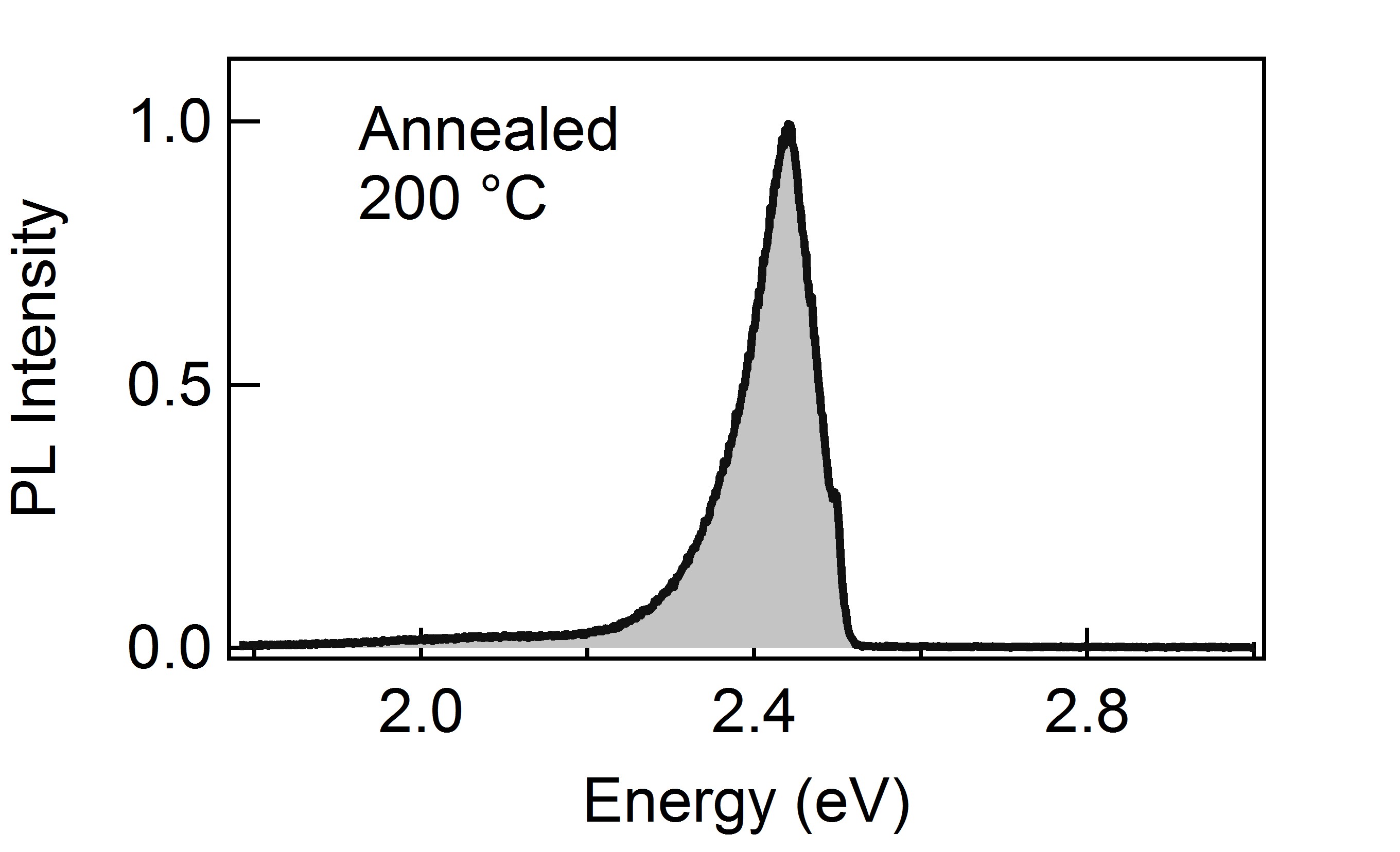}
\caption{Photoluminescence emitted in the visible spectral range from the MA200 sample at 10 K. The peak centered at 2.45 eV arises from carriers recombination in PbI2 inclusions.}
\label{7M}
\end{figure}

After having discussed the drift-diffusion of the electrons in high quality crystals, we can address the important role that sample degradation has on the carriers motion. The data of the as-grown crystal are compared with the ones of the samples annealed in air at 100 $^\circ$C (MA100) and 200 $^\circ$C (MA200) for 30 minutes. The MAPbI3 single crystal is very sensitive to humidity, illumination and annealing temperature. Figure \ref{6M}A shows photoluminescence on the as-grown and annealed samples. Upon increasing the annealing temperature, a new photoluminescence peak, arising from trapped states, develops $\cong60$ meV below the bandgap value.  We report in \ref{6M}B,C the photoelectron intensity map acquired on annealed samples as a function of pump-probe delay. The MA100 and the as-grown sample display similar behavior on the short timescale ($\cong 1$ ps). However the 2PPE map of MA100 (Fig. \ref{6M}B) and  MA200 (Fig. \ref{6M}C) holds a remnant signal up to 400 ps. We evince that shallow traps introduced by annealing lead to a partial localization of the electrons. Note in Fig. \ref{6M}D that the long timescale evolution of integrated 2PPE signal deviates from the diffusive $1/\sqrt{t}$ decay only in the case of annealed samples. The onset of the trapping strongly depends on the quality of the crystal. In MA100 the localization takes place on the timescale of few picoseconds whereas it falls below the picosecond in highly degraded MA200. Probably the defect density of MA200 is so high that localization takes place as soon as the electrons cool down below the mobility edge of the traps. The localization landscape of conduction electrons is related to the compositional disorder of degraded samples \cite{Jemli}. According to Xie \textit{et al.}, an annealing to 150 $^\circ$C leads to a shallow distribution of I and Pb components in the MAPbI3 thin film \cite{Xie}. Furthermore, Deretzis \textit{et al.} reported that thermodynamic degradation above 150 $^\circ$C can induce a partial conversion of MAPbI3 in to PbI$_2$ \cite{Deretzis}. The low temperature photoluminescence measurements in Fig. \ref{7M} confirms that MA200 contains large inclusions of PbI$_2$.

\section{Radiative and surface recombination}

\begin{figure}
\includegraphics[width=0.68\columnwidth]{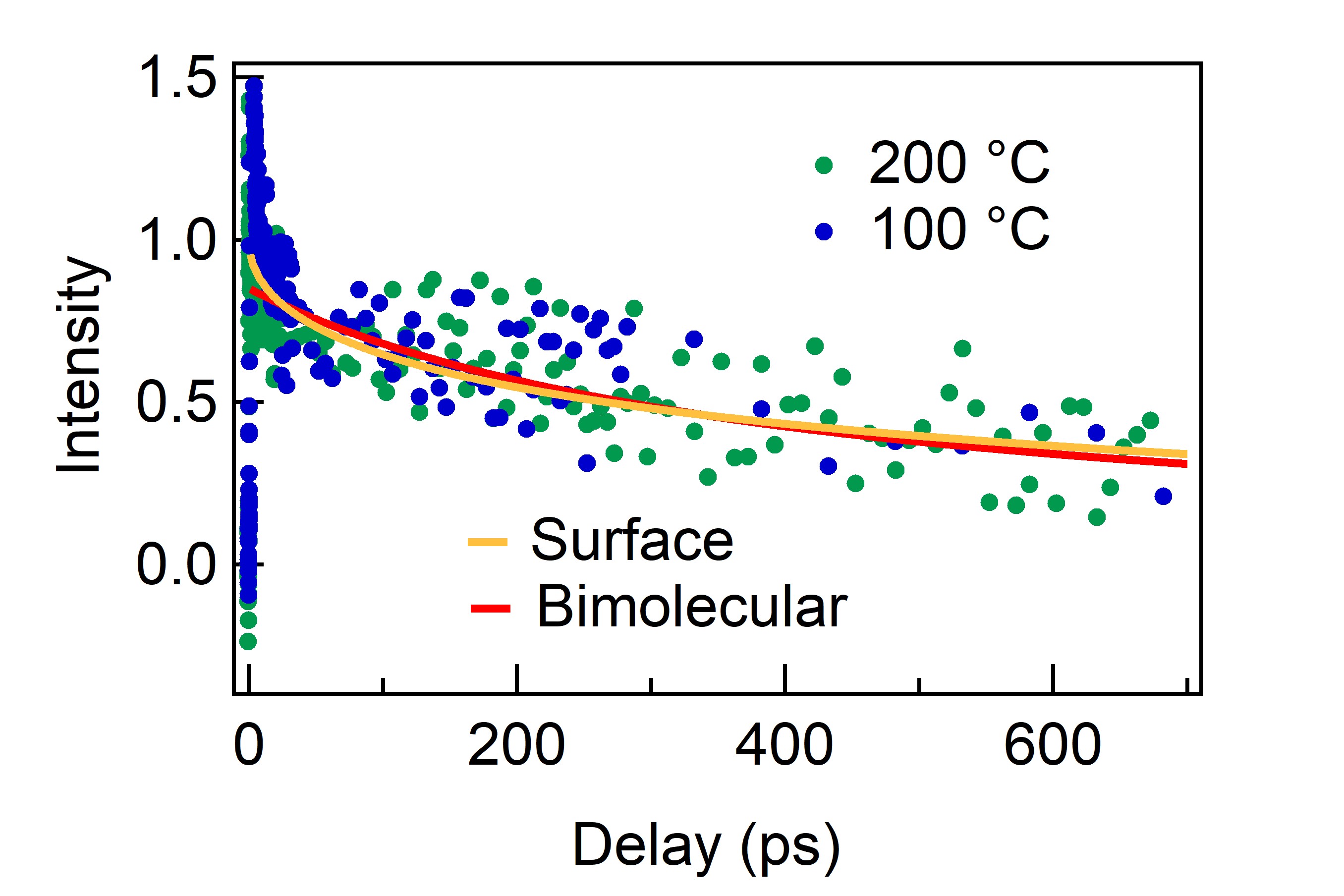}
\caption{Temporal evolution of the integrated 2PPE signal in the sample annealed at $100 ^\circ$C (blue marks) and $200 ^\circ$C (green marks). The red line is a fitting curve with bimolecular recombination rate $\gamma=4\pm1\times 10^{-10}$ cm$^3$/s whereas the yellow line (which is almost indistinguishable from the red line) is the fit obtained by equation \ref{eq2} with $S=4000$ cm/s and $D_T=0.05$ cm$^2$/s.}
\label{8M}
\end{figure}

Remarkably, the long lasting presence of electrons at the surface of MA100 and MA200 indicates a long recombination time of the photoexcited carriers. It has been shown that direct recombination between electrons and holes \cite{Herz_Reco} becomes dominant at photoexcitation density $\sim 10^8$ cm$^{-3}$. We display in Fig. \ref{8M} the fitting curve obtained by a kinetic equation with bimolecular recombination rate $\gamma=4\pm1\times 10^{-10}$ cm$^3$/s. The extracted value is in good agreement with previous reports and defies the Langevin prediction by four orders of magnitude \cite{Herz_Reco}. Wehrenfennig \emph{et al.} proposed that electrons and holes are prefentially localized in spatially distinct regions of the disordered landscape. Accordingly, density functional calculations on MaPbI3 predict that valence band maxima consist of $6s$- and $5p$ orbitals of lead and iodine, respectively, while conduction band minima mostly incorporate $6p-$orbitals of lead \cite{Nakao}.

Next, we exploit the high surface sensitivity of 2PPE to question the surface recombination velocity of samples cleaved in ultra high vacuum conditions. By following the literature \cite{Beard_Reco,Hoffman}, we model the density of the electrons at the surface as 

\begin{equation}
n(0,t)\propto \frac{\alpha D_T w\left(\alpha\sqrt{D_Tt}\right)-Sw\left(S\sqrt{\frac{t}{D_T}}\right)}{\alpha D_T-S},
\label{eq2}
\end{equation}

where $D_T$ is the diffusion constant of trapped electrons and $S$ is the velocity of surface recombination. Figure \ref{8M} compares the long timescale dynamics measured in the MA100 and MA200 crystals with the fit obtained for $S=4000$ cm/s and $D_T=0.05$ cm$^2$/s. Such $S$ value is strictly an upper bound because: i) we chose the $D_T$ value that provides a good fit and maximize $S$, ii) we implicitly assume that the observed decay is ruled by surface recombination instead of bimolecular recombination. The upper bound of $S$ is not far from the surface recombination velocity estimated in other hybrid perovskites \cite{Beard_Reco, Wu}. It should be outlined that $S<4000$ cm/s is two or three orders of magnitude lower than the surface recombination velocity of most unpassivated semiconductors \cite{Schmutten, Riffe}. Accordingly, the electronic structure calculations do not predict mid-gap states at MaPbI3 surfaces with thermodynamical stability\cite{Haruyama}. In operating devices, this asset favors the efficient extraction or injection of carriers at the electrical contacts with the MAPbI3 layer.

\section{Conclusions}

To conclude, we characterized the dynamics of excited electrons at the surface of MAPbI3. Our data provide a direct visualization of the electronic cooling at early delay times. It follows that photoexcited carriers thermalize on a subpicosecond timescale, presumably because of the coupling to the vibrations of organic cations. In the as-grown crystal, the electrons dynamics is ruled by diffusion at long timescales. Most likely, an additional drift due to built-in fields sets in at early delay.

We intentionally induce compositional disorder in some MAPbI3 crystals by thermal annealing in atmospheric conditions. As a result, the photoexcited electrons are localized by shallow traps within few picoseconds. Such localization mechanism is consistent with the drop of photoconversion efficiency in aged cells. Finally we estimate the surface recombination velocity of MAPbI3 cleaved in ultra high vacuum. The upper bound obtained by our analysis is consistent with previous results and is several orders of magnitude lower than the values reported in many unpassivated semiconductors.

The project leading to this article has received funding from the European Union's Horizon 2020 research and innovation program under grant agreement No 687008 (GOTSolar). We also thanks dim-nanoK for funding under project 'PIED', the DIM-Oximore and the Ecole Polytechnique for funding under the project 'ECOGAN'.

\end{document}